\documentclass[9pt,preprint2,natbib209]{aastex}
\shorttitle{Dust Formation in Common Envelopes} \shortauthors{L\"{u}
et al.}

\begin{document}


\title{Dust Formation in the Ejecta of Common Envelope Systems}
\author{Guoliang L\"{u}\altaffilmark{1}, Chunhua Zhu\altaffilmark{1},
 Philipp Podsiadlowski\altaffilmark{2}}
\email{$^\dagger$guolianglv@gmail.com}
\altaffiltext{1}{School of Physical Science and Technology, Xinjiang
University, Urumqi, 830046, China.} \altaffiltext{2}{Department of
Astronomy, Oxford University, Oxford OX1
3RH.}
%\altaffiltext{3}{Zentrum f\"{u}r Astronomie, Institut f\"{u}r
%Theoretische Astrophysik, Universit\"{a}t Heidelberg,
%Albert-\"{U}berle-Str. 2, D-69120 Heidelberg, Germany.}

%\date{}

%\pagerange{\pageref{firstpage}--\pageref{lastpage}} \pubyear{2007}

%\maketitle

%\label{firstpage}

\begin{abstract}
The material that is ejected in a common-envelope (CE) phase in a
close binary system provides an ideal environment for dust
formation. By constructing a simple toy model to describe the
evolution of the density and the temperature of CE ejecta and using
the \emph{AGBDUST } code to model dust formation, we show that dust
can form efficiently in this environment.  The actual dust masses
produced in the CE ejecta depend strongly on their temperature and
density evolution. We estimate the total dust masses produced by CE
evolution by means of a population synthesis code and show that, compared
to dust production in AGB stars, the dust produced in CE ejecta may be
quite significant and could even dominate under certain circumstances.

\end{abstract}


\section{Introduction}
Dust is one of the important ingredients of the interstellar medium
(ISM). It plays a central role in the astrophysics of the ISM, from
the thermodynamics and chemistry of the gas to the dynamics of star
formation. It affects the thermal and chemical balance of the ISM by
reprocessing the radiative output from stars, providing photoelectrons
which heat gas, and depleting the gas of refractory elements, which are
important cooling agents in the ISM \citep[e.g.,][]{Nozawa2007}. In
addition, dust changes the spectra of galaxies: radiation at short
wavelengths is attenuated, and energy is radiated in the infrared.
\cite{Bernstein2002} estimated that 30\% or more of the energy emitted
as starlight in the Universe is re-radiated by dust in the infrared.

From a theoretical point of view, the formation and growth of dust
grains is still a widely unsolved problem \citep[e.g.,][]{Gail1999,Gall2011}. 
There are numerous efforts
under way trying to understand the processes 
involved\cite[e.g.,][]{Gail1999,Todini2001,Draine2009}. Up to now,
people have assumed that dust mainly originates from the stellar winds of
asymptotic giant branch (AGB) stars and supernova (SN) ejecta.  Gail
and his collaborators investigated the formation and growth of dust
grains produced by AGB stars
\citep[e.g.,][]{Gail1999,Ferrarotti2006,Gail2009}, while  \cite{Todini2001},
\cite{Nozawa2003} and \cite{Bianchi2007}, among others, have studied
the condensation and survival of dust grains in SN ejecta.

Common envelopes (CE) form as a result of dynamical timescale mass
exchange in close binaries and play an essential role in their
evolution \citep[see, e.g.,][]{Paczynski1976,Iben1993}. In most cases,
CE evolution involves a giant star transferring matter to a normal or
a degenerate star on a dynamical timescale. The giant envelope
overfills the Roche lobes of both stars and engulfs the giant core and
its companion. During the CE phase, orbital energy is transferred to
the CE via dynamical friction between the orbiting components and the
non-corotating CE. If enough orbital energy is deposited in the CE
before the components merge, the whole envelope can be ejected on a
dynamical timescale. The CE ejecta then rapidly expand. According to
the classical theory of nucleation \citep{Feder1966}, the saturation
pressure in most cases falls more rapidly than the pressure of the
vapour, and the vapour becomes supersaturated when saturated vapour
expands adiabatically. The formation of dust grains from the gas phase
can occur from vapour in a supersaturated state. Therefore, CE ejecta
provide a potentially important environment for the formation of dust
grains. To our knowledge, there is no investigation of dust formation
in CE ejecta to date.

In this paper, we investigate dust grain formation and growth in CE
ejecta. \S 2 describes the model for the CE ejecta, and \S 3 gives the
input parameters for the dust modeling. The main results are presented
in \S 4 and discussed in detail in \S 5.  In \S 6 we estimate the total
dust masses produced using population synthesis modeling, and \S 7
summarizes our main conclusions.

\section{The Mass Density and Temperature of Common Envelope Ejecta}

In general, a CE system contains a gainer (a normal star or compact
object) and a donor composed of a giant envelope and a giant core. The
CE ejecta are formed from the ejected giant envelope. In general, dust
formation and growth is determined by the density and temperature
evolution of the CE ejecta.  Unfortunately, our understanding of CE
evolution is still rather poor despite numerous efforts to improve our
understanding \citep[e.g.,][]{Ricker2008,Ge2010,Deloye2010}. To
first-order approximation, we assume that the whole giant envelope is
ejected as CE ejecta on a dynamical timescale. We use the EZ code
\citep{Paxton2004}, derived from Eggleton's STARS code
\citep{Eggleton1971}, to simulate the giant's structure and
evolution. In this work, `giant envelope' refers to the region in the
star in which the hydrogen abundance (by mass) is larger than 0.5.  In
the EZ code, the giant envelope is divided into $\sim 100$ zones. The
mass density and the temperature in every zone are noted by $\rho_0$
and $T_0$, respectively. These determine the initial conditions for the
CE ejecta.

After the giant envelope has been ejected, it begins to expand, and
the density and temperature start to decrease.  \\

\noindent
(i){\it The Evolution of Mass Density --} The mass density $\rho$ is
represented by $\rho={\dot{M}_{\rm ej}}/({4\pi R^2 V})$, where
$\dot{M}_{\rm ej}$ and $R$ are the mass-ejection rate and the radial
distance of the ejected matter, respectively, and $V$ is the velocity
of the ejected matter. According to \cite{Harper1996}, the main
characteristic of the cool winds of evolved K and early M giants is
that their terminal velocities is lower than the surface escape
velocity, typically, $V_{\infty}\leq \frac{1}{2}V_{\rm
  esc}=\frac{1}{2}\sqrt{\frac{2GM_{\rm t}}{R_{\rm s}}}$, where $V_{\rm
  esc}$ is the escape velocity at the stellar surface, $R_{\rm s}$ is
the stellar radius and $G$ is the gravitational constant; $M_{\rm
  t}=M_{\rm gainer}+M_{\rm donor}$, where $M_{\rm gainer}$ is the mass
of the gainer and $M_{\rm donor}$ is the mass of the donor.  If the
velocity of the CE ejecta at infinity is similar to the terminal
velocity of the cool wind from the red giant, energy conservation can
be used to determine the velocity $V$ as a function of distance $R$
from the giant from $\frac{1}{2}(V^2 - V_{\infty}^2) = G M_{\rm t}/R$.
Considering the stellar structure of a red giant, we assume, for
simplicity, that $V$ approximately equals $\sqrt{\frac{2GM_{\rm
      t}}{R}}$, that is, $V$ is the escape velocity at $R$. It is
difficult to determine $\dot{M}_{\rm ej}$ for CE ejecta. Since we
assume that the giant envelope is ejected on a dynamical timescale, we
assume for each layer within the CE ejecta that $\dot{M}_{\rm
  ej}=\rho_04\pi R_0^2 V$, where $R_0$ is the initial distance of this
layer to the stellar center.  Therefore, the evolution of the density
in each layer can be represented by
\begin{equation}
\rho=\left(\frac{R_0}{R}\right)^{3/2}\rho_0.\label{eq:rho}
\end{equation}

(ii){\it The Evolution of Temperature --}
The expansion of the CE ejecta leads to the cooling of the gas. At the
beginning, most of the hydrogen atoms in the giant envelope are fully ionized
except for hydrogen near the giant's photosphere.
%The internal energy involves the ionization energy of H, the
%dissociation energy of H$_2$, and the basic $3/2 k T$ for a simple
%perfect gas. The internal energy can partly be used to drive the CE
%\citep{Han1995a}. However, it is difficult to accurately describe
%the temperature's evolution of CE ejecta. For simplicity, we divide
%CE ejecta into two zones. In the first zone, there is a large amount
%of ionized hydrogen.
As the temperature decreases, hydrogen recombines, gradually turning
ionized hydrogen nuclei into hydrogen atoms and releasing the recombination
energy in the process. Here, to first-order approximation, we only consider
the recombination of hydrogen. The released
energy can partly be absorbed by the CE ejecta and be used to drive
the CE expansion further \citep{Han1995a} or be lost from the CE ejecta (e.g.
by radiation). However, it is difficult to accurately describe the
efficiency of this process. For simplicity, we
introduce a parameter $\gamma$ to describe the temperature
evolution in each layer:
\begin{equation}
T=\left(\frac{R_0}{R}\right)^{\gamma}T_0, \label{eq:1zone}
\end{equation}
where $T_0$ is the initial temperature. The degree of ionization of
the hydrogen atoms is determined by the Saha equation
\begin{equation}
\frac{n_{+}n_{\rm e}}{n-n_{\rm e}}=\frac{2}{\sqrt{{ h^2}{2\pi
m_{\rm e}{ k_{\rm B}}T}}}
\exp\left({\frac{-\varepsilon}{{ k_{\rm B}}T}}\right),
\end{equation}
where $n$, $n_{+}$ and $n_{\rm e}$ are the density of hydrogen atoms,
hydrogen ions and electrons, respectively.  In the case of pure
hydrogen, $n_{+}=n_{\rm e}$.  The constants $h$, $k_{\rm B}$ and
$m_{\rm e}$ are the Planck constant, the Boltzmann constant and the
mass of the electron, respectively.  The ionization energy of hydrogen
is given by $\varepsilon$ and equals 13.6\,eV. The temperature $T$ is
given by Eq.~(\ref{eq:1zone}).  As the temperature decreases, the
degree of ionization of hydrogen, ${n_+}/{n}$, also decreases.  In
this work, we take ${n_+}/{n}=1\%$ as the boundary between the regions
where hydrogen is fully ionized and where it is partially ionized or
atomic. The distance of this boundary from $R_0$ is denoted as $R_{\rm
  cr}$, and the temperature of the CE ejecta at $R_{\rm cr}$ is
$T_{\rm cr}$.  Once most of the ionized hydrogen has recombined, the
gas in the CE ejecta will be similar to the gas at the surface of the
red giant.  Following \cite{Gail1999}, we refer to
\cite{Lucy1971,Lucy1976} to calculate the temperature evolution from
\begin{equation}
T^4=\frac{1}{2}T^4_{\rm cr}\left(1-\sqrt{1-\frac{R^2_{\rm
cr}}{R^2}}+\frac{3}{2}\tau_{\rm L}\right), \label{eq:nate}
\end{equation}
where $\tau_{\rm L}$ is defined by
\begin{equation}
\frac{{\rm d}\tau_{\rm L}}{{\rm d}r}=-\rho \kappa_{\rm
H}\frac{R^2_{\rm cr}}{R^2}.
\end{equation}
Here, $\kappa_{\rm H}$ is the flux averaged mass extinction
coefficient and is calculated by a simple
superposition of the extinction of the different dust species and
the gas (see details in \cite{Gail1999}).

In summary, the expansion of the CE ejecta is split into two zones. In the
inner zone ($R<R_{\rm cr}$), the evolution of the CE ejecta's temperature
is given by Eq.\ (\ref{eq:1zone}).  In this zone, the CE ejecta temperature
is high enough for hydrogen to be fully ionized so that dust cannot form.  In
the outer zone ($R>R_{\rm cr}$), the temperature's evolution is
described by Eq.\ (\ref{eq:nate}).  In this zone, the temperature of the CE
ejecta has dropped so that ultimately dust grains can form.

\section{A Dust Model for the Common Envelope Ejecta}
As mentioned in the last section, we assume that the structure of the
CE ejecta in the outer zone, where $R>R_{\rm cr}$, is similar to that
in the stellar wind from an AGB star. With the CE ejecta expanding,
its temperature and pressure decrease, and some tiny seed nuclei can
form in the cooling gas. However, the nucleation of seed nuclei from
the gas phase is a difficult problem. The \emph{AGBDUST} code does not
consider this problem and assumes that the seed nuclei already
exist. Condensation of dust can occur on the surface of the seed
nuclei. Then, dust starts to grow. \cite{Gail1999} and \cite{
  Ferrarotti2001,Ferrarotti2002,Ferrarotti2003,Ferrarotti2005} have
investigated the condensation and growth of dust grains in the stellar
wind from AGB stars using the \emph{AGBDUST} code. In this work, we
use the same code for dust formation in CE ejecta. In the
\emph{AGBDUST} code, the condensation and growth of dust grains is
affected by the following input parameters (other input parameters
that are not specifically mentioned are taken to have the default
values as in \cite{Ferrarotti2006}):\\
(i) \emph{Chemical composition --} The chemical composition has a
large effect on dust formation. In this paper, we investigate dust
formation in CE ejecta formed in binary systems in which giants are on
the first giant branch (FGB). Due to the first dredge-up, the chemical
abundances in the envelopes of FGBs star are different from the initial
abundances, which are taken from \cite{Anders1989} for the Sun. The
effects of the first dredge-up are a reduction of $^{12}$C by
approximately 30\% and no change in the $^{16}$O abundance at the
stellar surface \citep{Iben1983}. For Fe, Mg and Si, which are some of
the key elements for dust formation, the abundances are not changed
substantially. Therefore, the abundance ratio of carbon to oxygen by
number (C/O) in the CE ejecta is approximately 0.4. According to
\cite{Gail1999}, the most abundant dust species formed in the
circumstellar matter (where C/O$<1$) are olivine- and pyroxene-type
silicate grains and iron grains.\\
(ii) \emph{Temperature --} The temperature of the CE ejecta is determined by
Eqs.\ (\ref{eq:1zone}) and (\ref{eq:nate}) and is affected by the
parameter $\gamma$, a parameter that is quite uncertain.
Based on the results of model calculations by
\cite{Fransson1989}, \cite{Kozasa1989} adopted an adiabatic index
$\gamma_{\rm ad}=1.25$ for the early stage of SN explosions. For an
adiabatically expanding perfect gas, $\rho T^{\frac{1}{1-\gamma_{\rm
      ad}}}={\rm constant}$. We assume that the temperature evolution
of the CE ejecta in the inner zone is similar to that in the early
stage of a SN explosion. This implies $\gamma=0.375$ in
Eq. (\ref{eq:1zone}). In this work, in order to check the effects
of this parameter on the dust-formation
process, we simulate cases with $\gamma=0.2$, 0.3 and 0.4.  \\
(iii) \emph{The mass density --} The mass density is determined by
Eq.(\ref{eq:rho}). The density profile will also be affected by the
geometry of the CE ejecta.  \cite{Sandquist1998} showed that the mass
loss in the orbital plane is about 5 times larger than in the polar
direction. It is beyond the scope of this paper to simulate such more
realistic structures. Specifically, in this work we assume that the CE ejecta
expand spherically.\\
(iv) \emph{The stellar luminosity and mass --} The stellar
luminosity is taken to be the luminosity of the FGB star, and the mass
is the mass of the binary, including the FGB star and the degenerate
companion. However, their effects are weak because the region of dust
formation is far away from the binary system.

\section{Results}
\begin{figure*}
\begin{tabular}{cc}
\includegraphics[totalheight=3in,width=2.in,angle=-90]{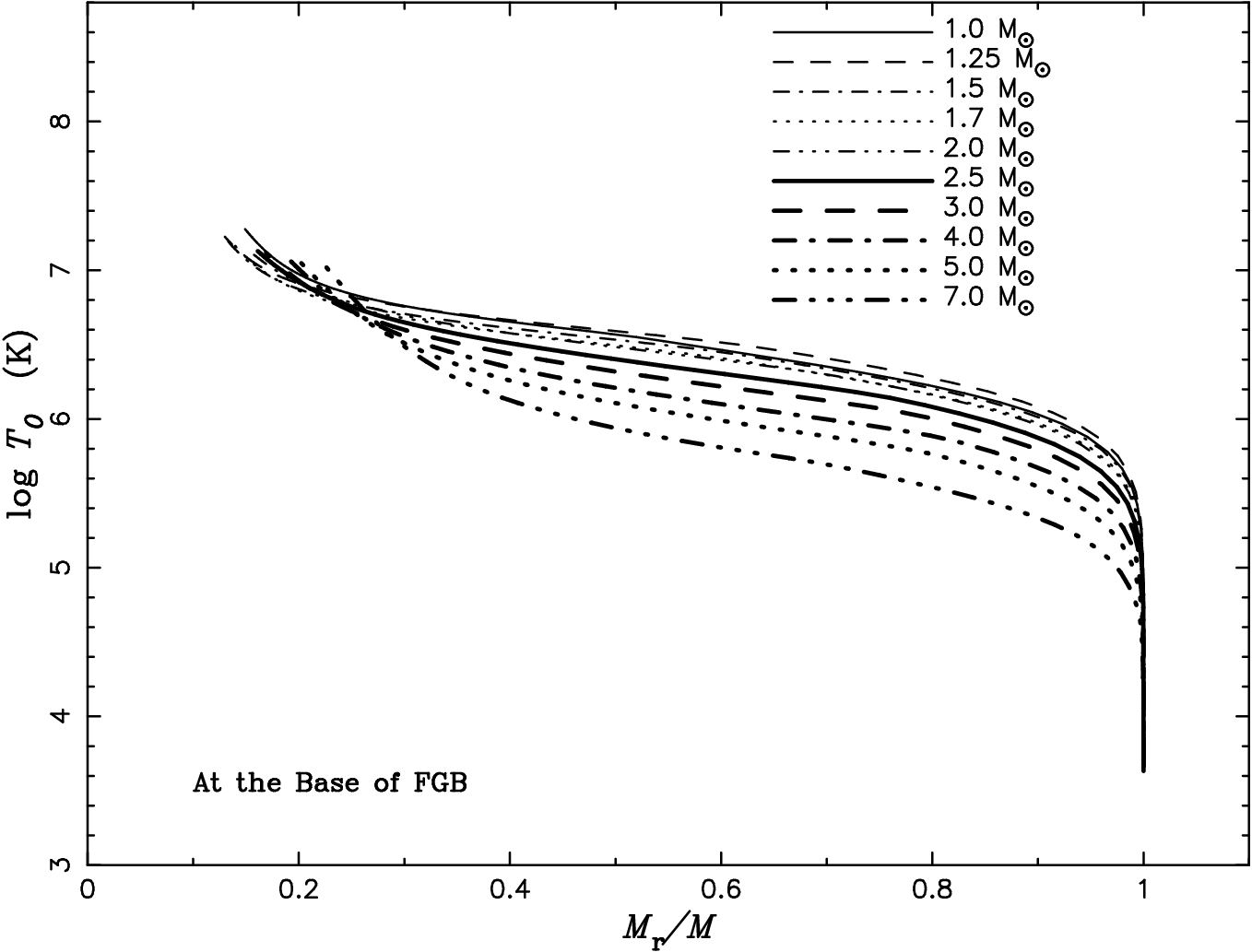}
&\includegraphics[totalheight=3in,width=2.in,angle=-90]{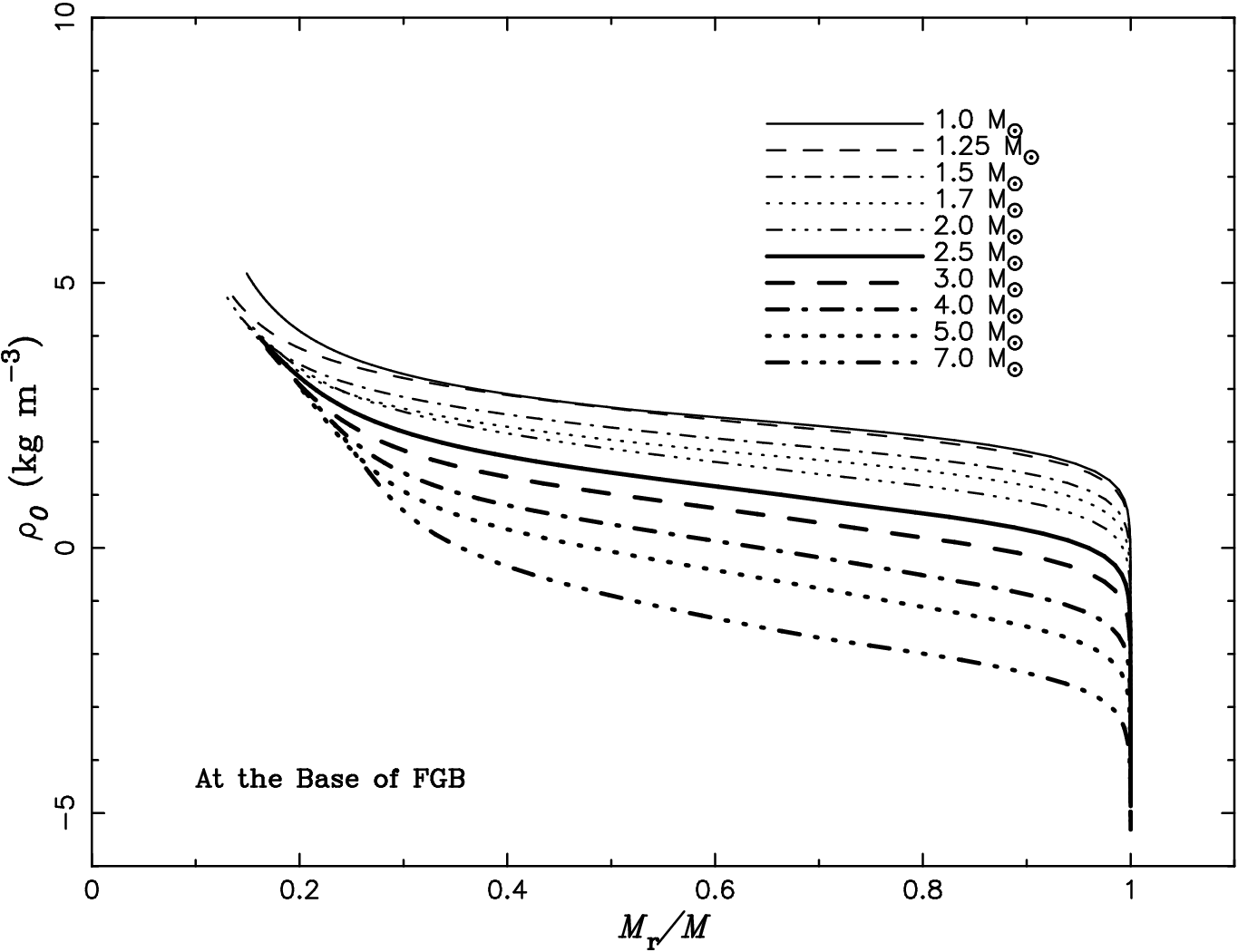}\\
\includegraphics[totalheight=3in,width=2.in,angle=-90]{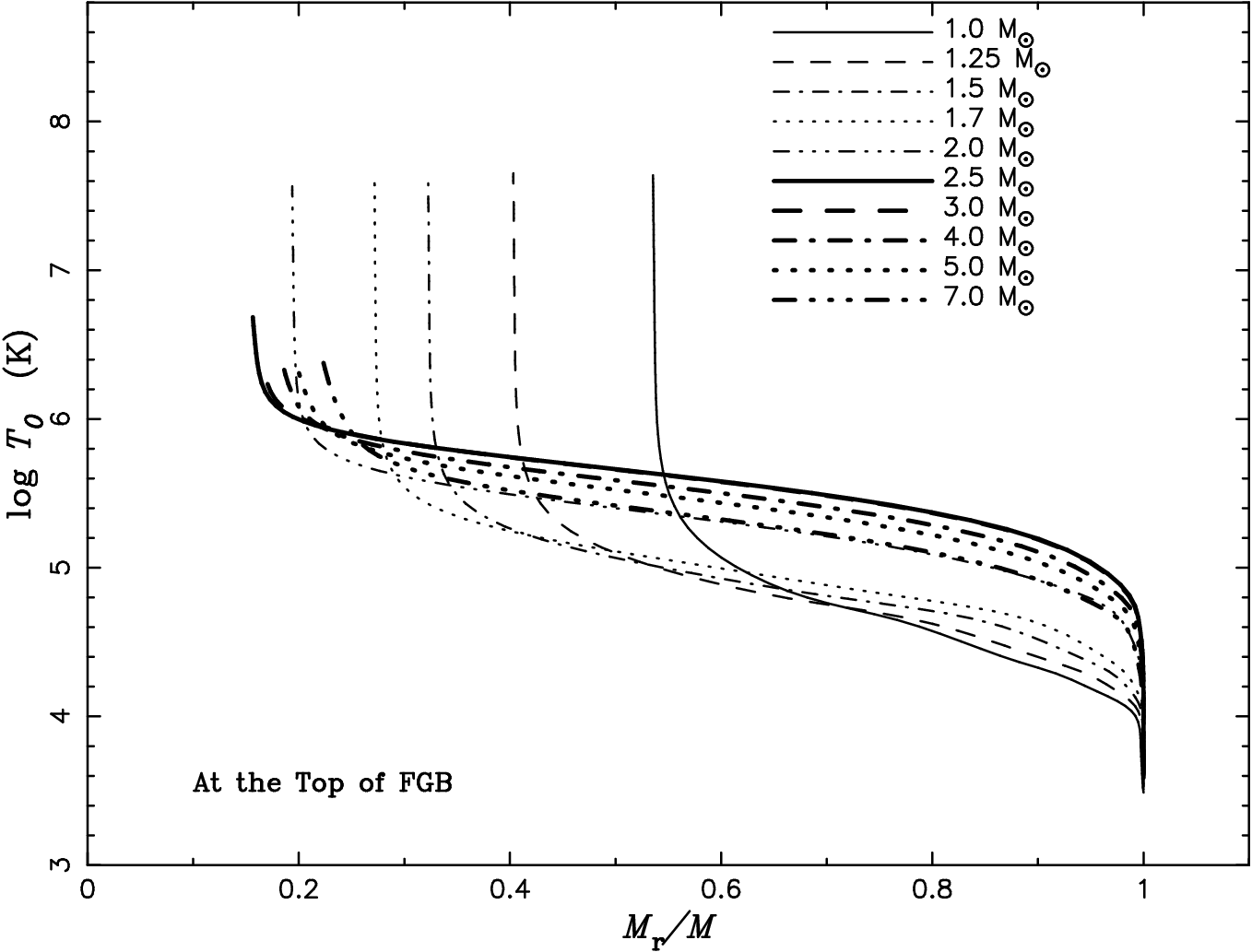}
&\includegraphics[totalheight=3in,width=2.in,angle=-90]{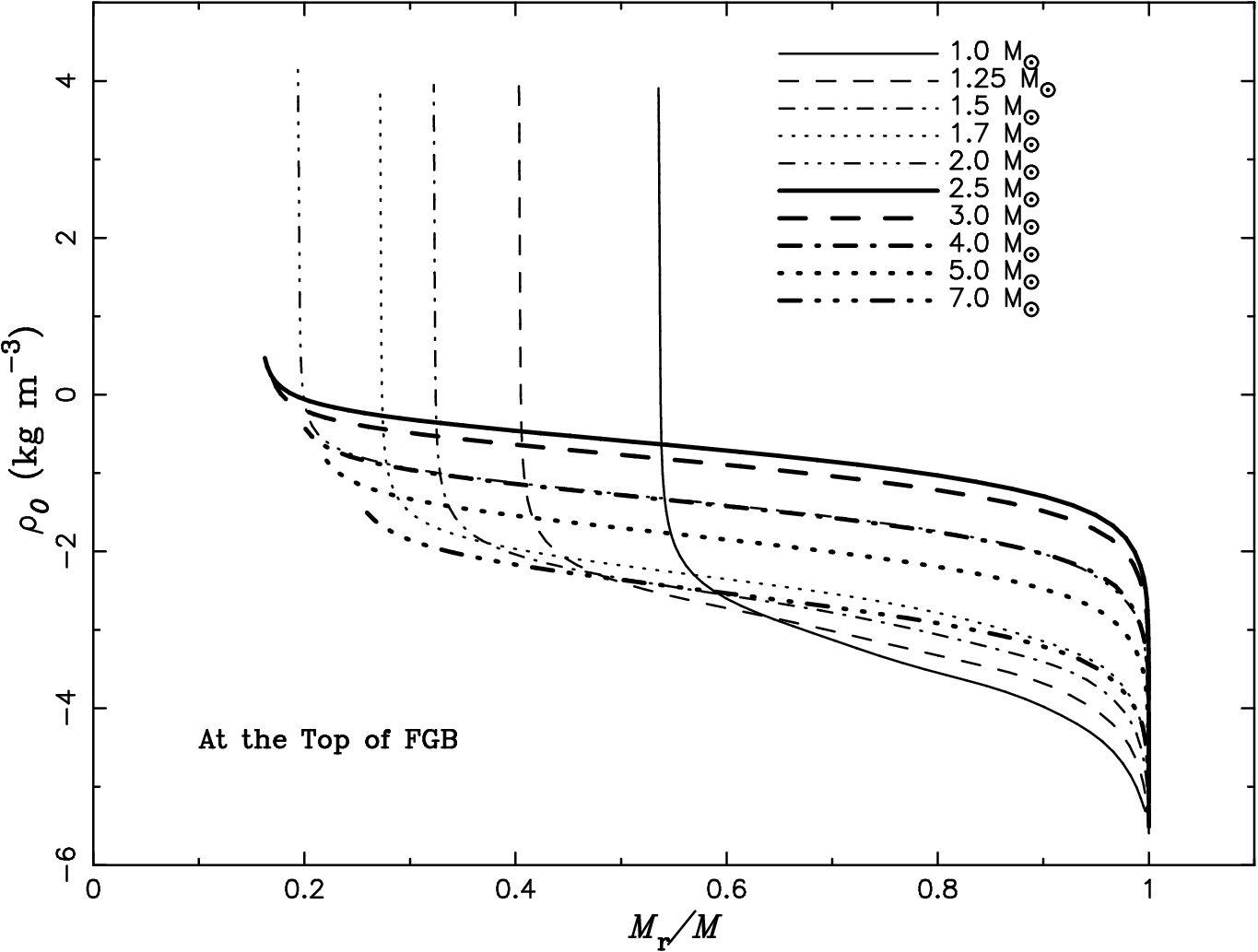}\\
\end{tabular}
\caption{The initial temperature and density profiles of the
envelopes of giants (at different evolutionary stages as indicated)
as a function of relative mass coordinate.}
\label{fig:initial}
\end{figure*}

Using the EZ code, we simulate the evolution of 10 stars with initial
masses of 1.0, 1.25, 1.5, 1.7 2.0, 2.5, 3.0, 4.0, 5.0 and 7.0
$M_\odot$, respectively. Their companions are assumed to be degenerate
stars of 1.0 $M_\odot$. We assume that they experience a CE phase
immediately when they overflow their Roche lobes on the FGB, ejecting
their whole envelopes.  The density and the temperature of the CE
ejecta depend on the parameter $\gamma$, and the initial temperature and
density profiles $T_0$ and $\rho_0$. $T_0$ and $\rho_0$ in every layer
are given by the EZ code and depend on the evolutionary state of the
giant.  In order to examine the effects of $T_0$ and $\rho_0$ on dust
formation, we calculate the dust masses produced in the CE ejecta
which originate from giants at the top of the FGB and at the base of
the FGB, respectively. Figure \ref{fig:initial} shows $T_0$ and
$\rho_0$ at every level of the giant envelope at the top of the FGB
and at the base of the FGB.

\begin{figure}
\includegraphics[totalheight=3.3in,width=3.0in,angle=-90]{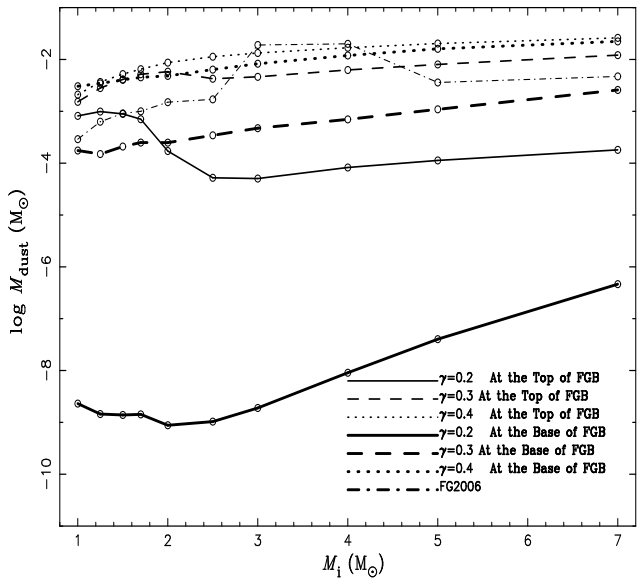}
\caption{Dust masses as a function of initial stellar mass. FG2006
refers the results in \cite{Ferrarotti2006}. The circles give
the dust masses in the models calculated. }\label{fig:qual}
\end{figure}

Figure \ref{fig:qual} shows the amount of dust produced (in $M_\odot$)
in the CE ejecta in these models for different values of $\gamma$. As
the figure demonstrates, $\gamma$ and the evolutionary state of the giant
dramatically affect the dust masses produced.  In particular, varying
$\gamma$ from 0.2 to 0.4 introduces an uncertainty for the dust masses
of up to a factor of $\sim10^7$.  Different $T_0$ and $\rho_0$
profiles, i.e. varying the evolutionary state from the base of the FGB
to the top of the FGB, introduce a variation of up to factor of
$\sim10^6$, but this depends strongly on the mass of the giant.

\begin{figure}
\includegraphics[totalheight=3.3in,width=3.0in,angle=-90]{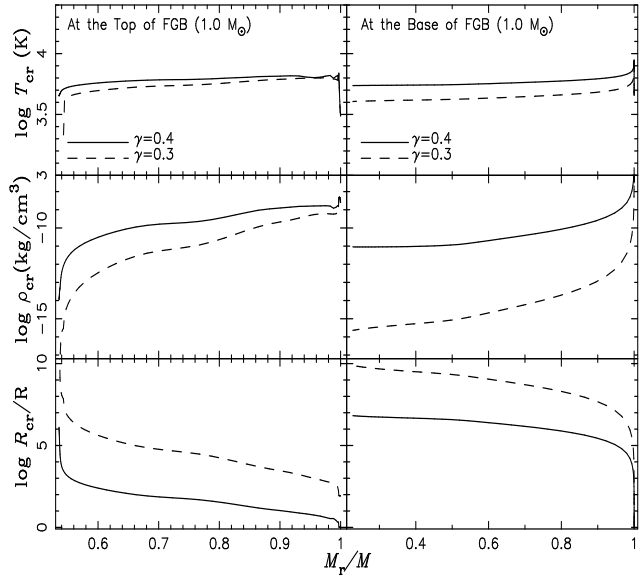}
\caption{Temperature, density and relative distance  as a function
of relative mass coordinate of the CE ejecta at $R_{\rm cr}$. The
left panels are for the CE ejecta produced by a star with $1 M_\odot$
at the top of the FGB. The right panels are for the CE ejecta produced by
a star with $1 M_\odot$ at the base of the FGB. Solid and dashed curves
represent $\gamma=0.4$ and $\gamma=$0.325, respectively.
}\label{fig:tempcr}
\end{figure}

The dust masses mainly depend on the temperature and the mass density
of the CE ejecta.  In order to show the effects of $\gamma$, $T_0$ and
$\rho_0$, we give the temperature, mass density and relative distance
of the CE ejecta produced by an FGB star of 1.0 $M_\odot$ at $R_{\rm
  cr}$ in Figure \ref{fig:tempcr}. Obviously, a higher $\gamma$
results in a smaller $R_{\rm cr}$ for the same $T_0$ and $\rho_0$, and
the CE ejecta have a higher density (see the right panels or the left
panels in Figure \ref{fig:tempcr}). According to \cite{Gail1999}, a
higher mass density results in a higher degree of dust condensation in
the stellar wind. Thus, in the simulation with a high $\gamma$, the CE
ejecta offer a very favourable environment for the formation and
growth of dust grains.  Similarly, a lower $T_0$ also results in a
smaller $R_{\rm cr}$ for the same $\gamma$ (see Figure
\ref{fig:tempcr}), and the CE ejecta have higher density at $R_{\rm
  cr}$.  Again dust grains can easily form and grow in the CE ejecta
in this case. In contrast, a low $\gamma$ and a high $T_0$ result in a
large $R_{\rm cr}$, and the mass density of the CE ejecta becomes too
low to lead to efficient dust production.

\begin{figure}
\includegraphics[totalheight=3.3in,width=3.0in,angle=-90]{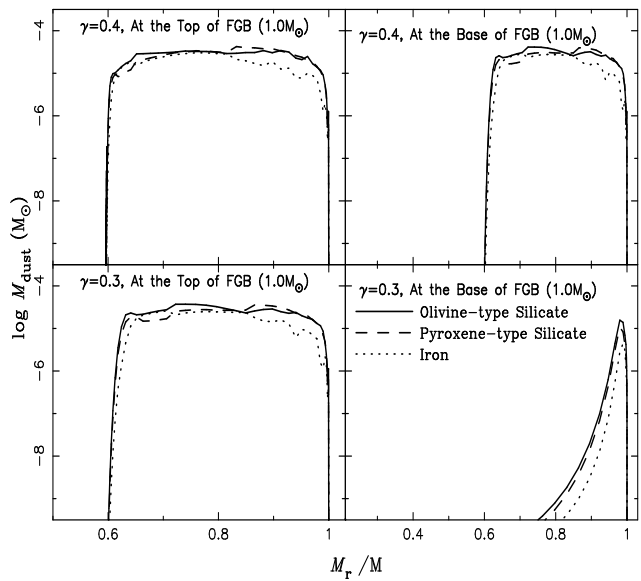}
\caption{Similar to Figure \ref{fig:tempcr}, but for the amounts of
olivine- and pyroxene-type silicate grains and iron grains produced
as a function of relative mass coordinate. }\label{fig:evequal}
\end{figure}

Beyond $R_{\rm cr}$, the CE ejecta enter the zone where dust may form.
Figure \ref{fig:evequal} gives the quantities of olivine- and
pyroxene-type silicate grains and iron grains produced by the CE ejecta.
Due to the small sticking efficiency for quartz \citep{Gail1999}, its
quantity in dust grains is negligible. In our simulations, the
silicates mainly consist of olivine and pyroxene whose proportion in
the dust grains is between $70-90\,$\%. The proportion of iron in
the dust grains is between $\sim 10 - 30\,$\%.  As Figure
\ref{fig:evequal} shows, for a small $\gamma$ and a high $T_0$,
dust is mainly produced by the CE ejecta close to the stellar
surface, while the whole CE ejecta can efficiently produce dust for
a high $\gamma$, and a low $T_0$ and $\rho_0$.

\cite{Ferrarotti2006} calculated the quantities of dust produced by
AGB stars. This is plotted as a dot-dashed curve in Figure
\ref{fig:qual}. This shows that, although dust production is comparatively
unimportant in the CE models with $\gamma=0.2$ at
the base of FGB, the dust quantities produced by the CE ejecta are
significant or may even dominate in other models with a higher $\gamma$.
There are two main reasons for this: \\
(i) \cite{Ferrarotti2006} concluded that the
dust condensation is efficient when AGB stars have high mass-loss
rates, higher than $2\times10^{-6}\, M_\odot\, {\rm yr}^{-1}$. For an AGB
wind with a high mass-loss rate of $1\times10^{-5}\, M_\odot\, {\rm
  yr}^{-1}$, the temperature and the mass density of dust-forming
zones are $\sim$ $1.3\times10^{3}$ K and $5.0\times10^{-14}{\rm
  g\ cm}^{-3}$, respectively. About 32\% of Si elements in an AGB
wind condensate into olivine-, pyroxene- and quartz-type dust, and
about 4\% of Fe elements condensate into iron grains. However, in the
simulation with $\gamma=0.4$ and at the top of FGB, the temperature
and the mass density of dust-forming zones in the CE ejecta are $\sim$
$1.1\times10^{3}$ K and $\sim$ $3.3\times10^{-11}{\rm g\ cm}^{-3}$,
respectively. The mass density of the CE ejecta in the dust-forming zone is
much higher than that in an AGB wind and hence is very favorable for dust
formation and growth. About 90\% of the Si elements in the CE ejecta
condensate into silicate grains, and about 50\% of the Fe elements in the
CE ejecta condensate into iron grains.  \\
(ii) Generally, about 7\% (for $M_{\rm i}=1.0\,M_\odot$) to 70\% (for
$M_{\rm i}=7.0\,M_\odot$) of the mass of an AGB star are lost at
mass-loss rates in excess of $2\times10^{-6}\, M_\odot\, {\rm
  yr}^{-1}$, and only this portion can efficiently produce
dust. However, as Figure \ref{fig:initial} shows, about 40\% (for
$M_{\rm i}=1.0\,M_\odot$) to 80\% (for $M_{\rm i}=7.0\,M_\odot$) of the
stellar mass can be ejected as a CE at the top of the FGB. Furthermore,
as the top-left panel in Figure \ref{fig:evequal} shows, dust can
efficiently form and grow throughout the CE ejecta in this case.

Therefore, dust formation and growth in CE ejecta with high
$\gamma$ is more efficient than in an AGB wind, and, under
these circumstances, CE ejecta may produce more dust than AGB winds.

In \cite{Ferrarotti2006}, due to the third dredge-up, the dust species
produced by AGB stars are olivine-, pyroxene- and quartz-type
silicates, iron, SiC and carbon dust grains. Stars with initial masses
between $\sim 2 - 4\,M_\odot$ can evolve into carbon stars. They
produce a large amount of carbon dust, which produces the bump in the
thick dot-dashed curve in Figure \ref{fig:qual}. In our work, because
the CE ejecta originate from FGB stars, the dust species produced in
the CE ejecta are olivine-, pyroxene- and quartz-type silicates and
iron grains. If an AGB star which has undergone the third dredge-up
experiences a CE phase, the C/O ratio in the CE ejecta may be higher
than 1. Then, other dust species (such as SiC and carbon dust grains)
can also form.

\begin{figure*}
\begin{tabular}{cc}
\includegraphics[totalheight=3.3in,width=3.0in,angle=-90]{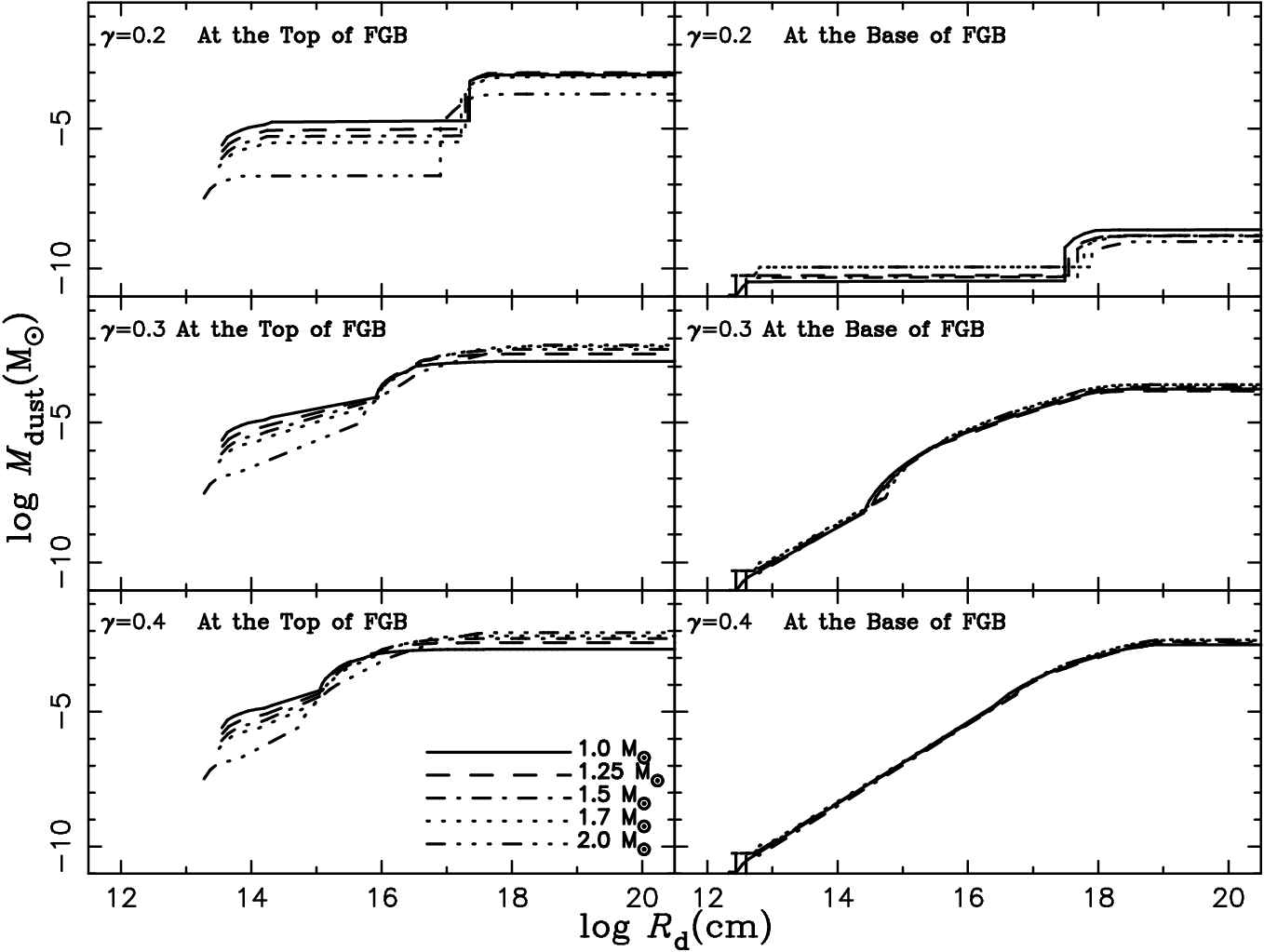}
&\includegraphics[totalheight=3.3in,width=3.0in,angle=-90]{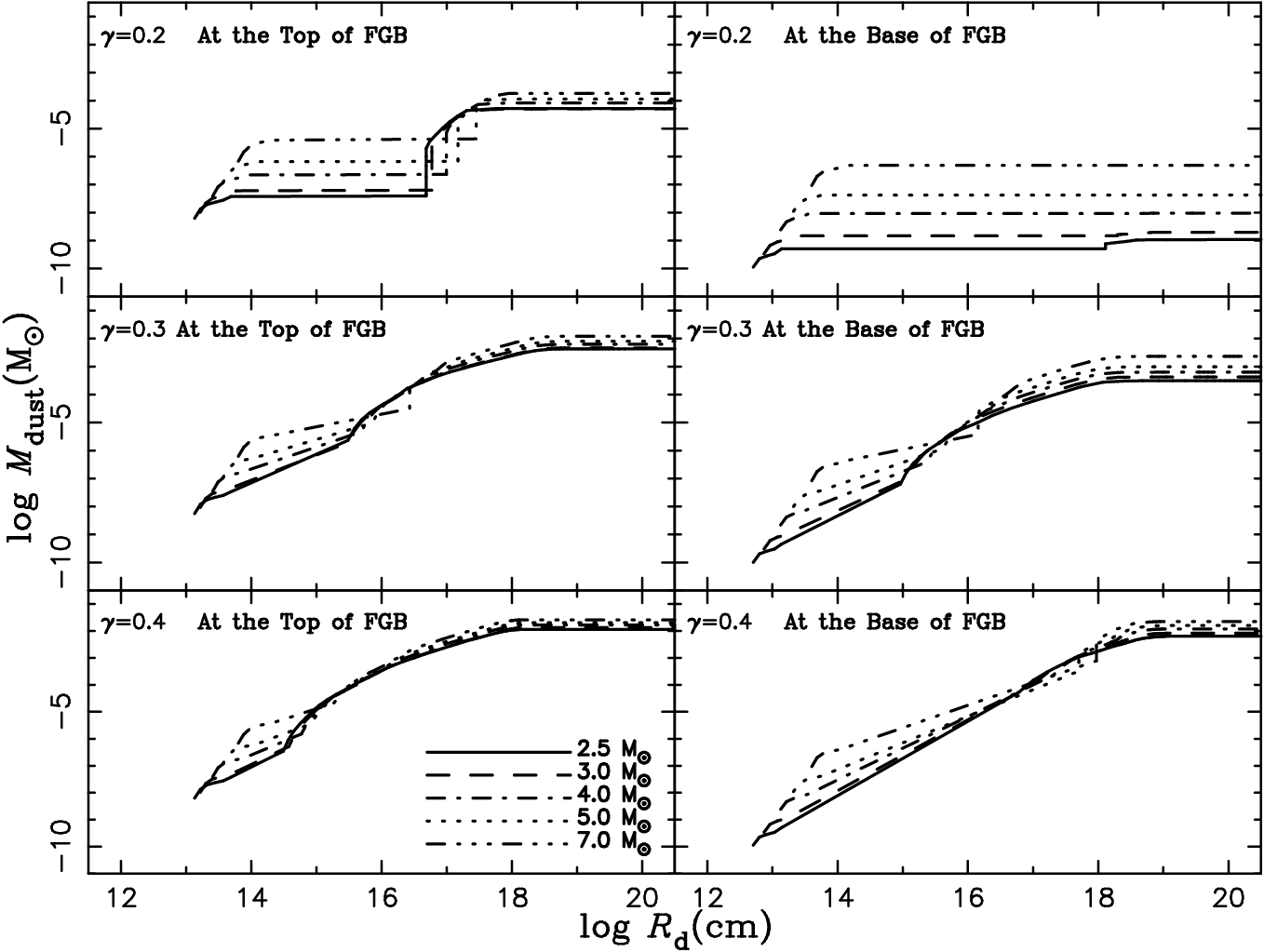}\\
\end{tabular}
\caption{The amounts of dust produced in the CE ejecta as a function
of the distance to the FGB star. The left panels are for FGB stars with masses
less than 2.0$\,M_\odot$, and the right panels are for FGB stars
with masses larger than 2.0$\,M_\odot$. } \label{fig:dist}
\end{figure*}

\section{Discussion}
As \S 4 suggests, under certain circumstances, CE ejecta can be very
efficient producers of dust. Hence, it should be possible to observe
large amounts of dust around post-CE systems.  \cite{Ferrarotti2006}
showed that the distance of dust formation from AGB stars is about
$10^{12} -10^{13}\,$cm. Figure \ref{fig:dist} shows the amount of dust
produced by CE ejection as a function of distance from the FGB
stars. The distance of dust formation in the CE ejecta is between
$\sim 10^{14} - 10^{18}\,$cm and is relatively far away from the FGB
star; this may make it difficult to observe the dust produced in CE
ejecta.

However, if there is an extreme event after the dust grains have
formed in the CE ejecta, for example, a SN in which a large amount of
energy is released, dust grains can be illuminated.  This may produce
light echoes, as have been observed in several SNe, by radiation
scattered by the dust near or along the line of sight.

For example, the progenitors of Type Ia SNe (SNe Ia) may have
experienced CE evolution.  From HST WFPC2 imaging,
\cite{Garnavich2001} proposed that SN 1998bu may have two echoes
caused by dust at $<10$ pc and 120 $\pm$ 15 pc away from the SN: the
inner echo is likely to be caused by dust from circumstellar material,
while the outer component is consistent with ISM dust. SN 1998bu is a
SN Ia. There are three other known SN Ia echoes, SNe 1991T, 1995E and
2006X. \cite{Wang2008} showed that these echoes have a wide range of
dust distances from $<10$ pc to $\sim$ 210 pc, which is consistent
with our results (see Figure \ref{fig:dist}). However, there are no
observational data for light echoes in the majority of SNe Ia. The
main reason is that it is very difficult to observe these light
echoes. In the single-degenerate model, there is typically a long time
delay between the CE phase and the supernova explosion of between
$10^8$ and $10^9$ years \citep{Han2004,Meng2009}. In the
double-degenerate model, the delay times often exceed $10^9$ years. In
contrast, the time it takes for material to move from the binary to
the zone of dust formation is only $10-100\,$yr. Dust grains formed in
the CE ejecta probably have significant outflow speeds.  It is
difficult to determine this outflow speed. If it is $\sim $ 10 km/s,
dust grains will have moved far away from the binary by the time of
the SN, and it will be difficult to observe light echoes. However, if
a degenerate WDs explode as SNe Ia within $10^7 - 10^8$\,yr after the
CE phase, it is possible to observe light echoes at distances of
$100-1000$\,pc (depending on the outflow velocity). The delay time in the single-degenerate model depends
on the mass of the companion of the WD. A delay time of $\sim
10^8$\,yr means that the companions of WDs have initial masses larger
than $\sim$ 6 $M_\odot$. Considering the condition for dynamical
stability for mass transfer, \cite{Han2004} suggested that the
main-sequence masses in the progenitors of SNe Ia cannot exceed $\sim
4.0\,M_\odot$. However, using observations of the evolution of the SN Ia
rate with redshift, \cite{Mannucci2006} suggested that SNe Ia have a
wide range of delay times, from $<10^8$ to $>10^{10}$ years. From a
theoretical point of view, assuming an aspherical stellar wind in
symbiotic stars, \cite{Lu2009} claimed that the initial masses of WD's
companions in progenitors of SNe Ia can exceed
6.0\,$M_\odot$. Similarly, \cite{Wang2009} argued that the delay times
of some SNe Ia from the helium-star channel are shorter than $10^8$
years. Therefore, it may be possible in principle to observe light
echoes in SNe Ia in which their progenitors are composed of WDs and
massive companions.

In one of the main channels for Type Ib SNe (core collapse supernovae
without hydrogen), the progenitors are believed to lose their hydrogen
envelopes in a CE phase \citep{Podsiadlowski1992}. The resulting naked
helium stars (which may appear as Wolf-Rayet stars) will explode after
the CE ejection within $\sim 10^4 - 10^6\,$yr (depending on the
evolutionary state at the beginning of mass transfer). Therefore, it
may be possible to observe light echos from dust grains in SNe Ib,
although their luminosities are much lower than those in SNe Ia. In
fact, dust grains have been observed in the case of SN2006jc, which
had a Wolf-Rayet star progenitor, though, in this case, it may be
more likely that the dust was produced in the SN ejecta themselves
\citep{Smith2008,Carlo2008,Nozawa2008}. Up to now, to our knowledge,
there is no direct observational evidence for dust grains in SNe Ib
formed by CE ejection.

\section{An Estimate of the Dust Produced by CE Ejecta via Population Synthesis}

CE evolution often occurs in close binary systems, and the CE ejecta
may provide an important contribution to the dust in the ISM.  Its
importance to the overall dust production can be estimated using
population synthesis modeling. Similar to the main case considered in
our study of symbiotic stars with WD components \citep{Lu2006}, we use
the initial mass-function of \citet{1979ApJS...41..513M} for the mass
of the primary components and a flat distribution of mass ratios
\citep{1989Ap.....30..323K,1994A&A...282..801G}. The distribution of
separations is determined by $\log a = 5X + 1$, where $X$ is a random
variable uniformly distributed in the range [0,1] and the separation
$a$ is in $R_\odot$.

During the CE phase, the binary experiences a dynamical spiral-in. It
is generally assumed that the orbital energy that is released by the
spiral-in process is used to expel the envelope of the donor with an
efficiency $\alpha_{\rm ce}$. In the theoretical calculation, the
dynamical spiral-in is affected by the combined parameter $\alpha_{\rm
  ce}\lambda_{\rm ce}$, where $\lambda_{\rm ce}$ parameterizes the
envelope structure of the donor. In this work, we take $\alpha_{\rm
  ce}\lambda_{\rm ce}=1.0$, i.e. assume that the CE ejection process
is very efficient.

Using the rapid binary-star evolution (BSE) code of \cite{Hurley2002},
we calculate the evolution of $10^6$ binary systems. About 20\,\% of
all binary systems undergo CE evolution and eject their envelopes. Of
these, about 51\,\% come from FGB stars, while the rest originate from
stars on the AGB.  For simplicity, we assume that all ejected CEs
originate from the stellar envelope either at the base of the FGB or
the top of the FGB. Usually, the envelopes of AGB stars have
temperature lower than those of FGB stars. Therefore, this assumption
may underestimate the amount of dust produced. According to Figure
\ref{fig:qual}, for given input parameters, the dust masses mainly
depend on the stellar mass at the beginning of the CE evolution. Using
one-dimensional linear interpolation of stellar masses, we estimate
the quantities of dust produced by CE evolution using this population
synthesis approach.  Similarly, using the results of
\cite{Ferrarotti2006}, we also estimate the quantities of dust
produced by AGB stars in a single starburst of $10^6$ single stars
through one-dimensional linear interpolation of stellar masses.

\begin{figure}
\includegraphics[totalheight=3.3in,width=3.0in,angle=-90]{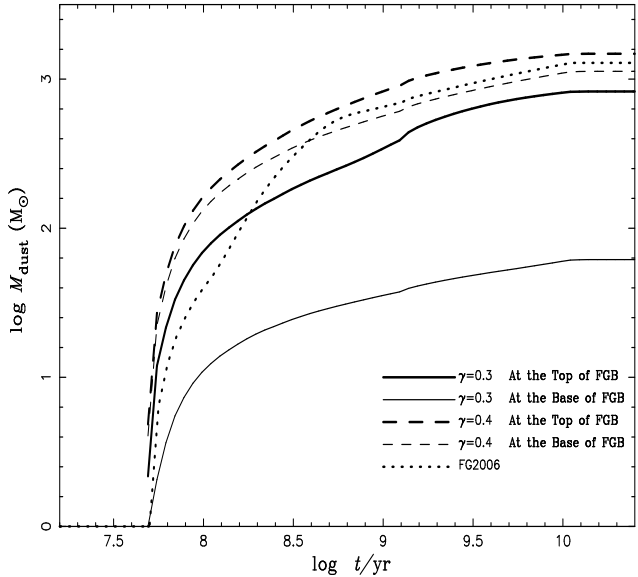}
\caption{The dust masses produced as a function of time
by a single starburst of
$10^6$ binary systems (thin and thick solid curves, thin and thick
dash curves) or $10^6$ single stars (dotted curve).
          }\label{fig:bse}
\end{figure}

Figure \ref{fig:bse} presents the dust masses produced by CE evolution
in a single starburst of $10^6$ binary systems or by $10^6$ single
stars. Compared with AGB stars, due to the high initial temperature and
slow temperature decrease, the quantities of dust produced by CE evolution
are negligible in the simulation with $\gamma=0.3$ at the base of the FGB.
However, CEs produce significant amounts of dust or
even dominate in the simulations with $\gamma=0.4$ and $\gamma=0.3$
at the top of the FGB. This demonstrates the potential importance
of CE evolution to the overall dust production in the ISM.

\begin{figure}
\includegraphics[totalheight=3.3in,width=3.0in,angle=-90]{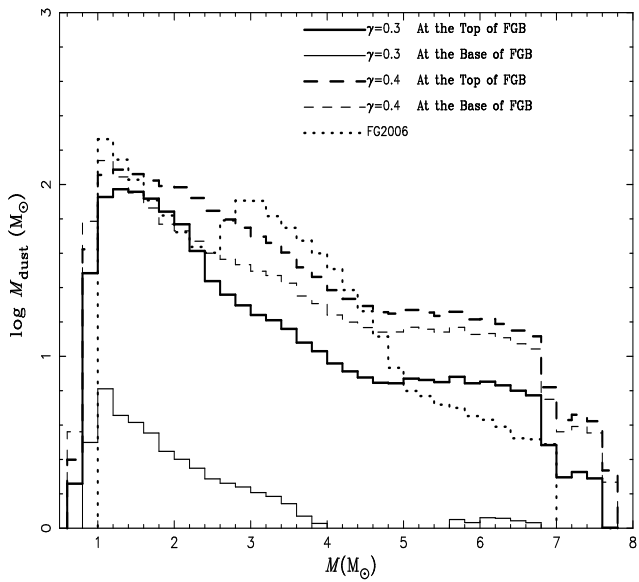}
\caption{The dust masses produced by stars of different masses in
$10^6$ binary systems (thin and thick solid curves, thin and thick
dash curves) or $10^6$ single stars (dotted curve).
          }\label{fig:psd}
\end{figure}

Figure \ref{fig:psd} shows how the dust masses produced by CEs or
single AGB stars depend on the initial stellar masses. There are two
peaks in the distribution of the dust quantities produced by AGB
stars. The left peak is a direct consequence of the initial mass
function that favours lower-mass stars, while the right peak
originates from stars with masses of $3-4 \,M_\odot$ which are
particularly efficient dust producers \citep{Ferrarotti2006}. The dust
produced by CEs mainly comes from the stars with masses of
$1-3\,M_\odot$ due to the initial mass function.

\section{Conclusions}
We have investigated the dust formation in CE ejecta constructing
simple models for the evolution of the CE ejecta and using the
\emph{AGBDUST} code to simulate dust formation.  The dust masses
produced by CE ejecta greatly depend on the input parameters. Compared
to the dust masses produced by AGB stars, they may be significant and
could even dominate under certain conditions.  This demonstrates that
the contribution of dust produced by CEs to the overall dust
production in the ISM needs to be taken into account.  The progenitors
of SNe Ia usually undergo CE evolution.  Due to the generally expected
long delay times of SNe Ia, it is difficult to observe the dust grains
formed in CE ejecta via light echoes, but it may be possible to find
light echoes produced by dust grains in SNe Ia in which their
progenitors are composed of WDs and massive companions. Due to the short
delay times, SNe Ib may be good candidates for observing dust grains
formed in CE ejecta. However, due to the still large uncertainties in
modeling the CE phase, our conclusions are very preliminary and
further work on its importance will be required.

\section*{Acknowledgments}
GL thanks H.-P. Gail for offering \emph{AGBDUST} code. This work was
supported by the National Natural Science Foundation of China under
Nos. 11063002 and 11163005, Foundation of Ministry of Education under No.
211198 and Foundation of Huoyingdong under No.
121107.
\bibliographystyle{apj}
\bibliography{lglapj}

%\begin{thebibliography}{99}
%\include{lgl.bib}
%\end{thebibliography}

\label{lastpage}

\end{document}